# Highly stable electronic properties of rippled antimonene under compressive deformation

Yujia Tian,[*,†] Devesh R. Kripalani,[*] Ming Xue,[†] Shaofan Li,[‡] and Kun Zhou[*,§]

[*]School of Mechanical and Aerospace Engineering, Nanyang Technological University, Singapore 639798, Singapore

[†]Infineon Technologies Asia Pacific Pte Ltd, Singapore 349282, Singapore

[‡]Department of Civil and Environmental Engineering, University of California, Berkeley, California 94720, USA

[§]Corresponding author. Email address: kzhou@ntu.edu.sg

Abstract

Antimonene has attracted much attention for its high carrier mobility and suitable band gap for electronic, optoelectronic, and even spintronic devices. To tailor its properties for such applications, strain engineering may be adopted. However, such two-dimensional (2D) crystals may prefer to exist in the rippled form because of the instability of long-range orders, and rippling has been shown to have a contrasting, significant impact on the electronic properties of various 2D materials, which complicates the tuning process. Hence, the effects of rippling on the electronic properties of antimonene under strain are herein investigated by comparing antimonene in its rippled and flat forms. Density functional theory calculations are performed to compute the structural and electronic parameters, where uniaxial compression of up to 7.5% is applied along the armchair and zigzag directions to study the anisotropic behavior of the material. Highly stable properties such as the work function and band gap are obtained for the rippled structures, where they are fully relaxed, regardless of the compression level, and these properties do not deviate much from those of the pristine structure under no strain. In contrast, various changes are observed in their flat counterparts. The mechanisms behind the different results are thoroughly explained by analyses of the density of states and structure geometry. The out-of-plane dipole moments of the rippled structures are also presented to give further insights into potential applications of rippled antimonene in sensors, actuators, triboelectric nanogenerators, *etc*. This work presents extensive data and thorough analysis on the effect of rippling on antimonene. The identification of optimal ripple amplitudes for which the electronic properties of the pristine condition can be recovered will be highly significant in guiding the rational design and architecture of antimonene-based devices.

# I. INTRODUCTION

The existence of two-dimensional (2D) crystals was questioned and debated over for a long time.[1-3] Ever since the successful isolation of graphene in 2004 that proves its stability,[4] other 2D materials such as antimonene,[5-9] phosphorene,[10-12] arsenene,[13-15] germanene,[16] monolayer $MoS_2$,[17-19] 2D ferroelectrics including SnS, SnSe, and SnTe[20], and 2D magnets[21] have sparked much interest. The reduced symmetries of these emerging materials lead to inhomogeneous electron distribution, different optical, valley, and spin responses, and properties including ferroelectricity, magnetism, and superconductivity.[22] These properties are different from those of the bulk and may offer more possibilities for next-generation electronic applications. However, the long-range order in 2D lattices will always be destroyed because of thermal fluctuation according to the Mermin–Wagner theorem,[2, 23] making ripple deformation ubiquitous in 2D free-standing sheets.

Rippling has been studied in detail for graphene,[24-26] monolayer $MoS_2$,[19] phosphorene,[12] *etc*. The band gap of graphene was found to be opened by corrugations with small curvatures because of the interruption of the mirror symmetries. Similarly, rippling was also reported to reduce the conductance of monolayer $MoS_2$. In direct contrast, the band gap of phosphorene decreases significantly when there are corrugations in its structure. Therefore, the effect of rippling on the electronic properties of 2D crystals cannot be generalized and is inherently material-dependent. To fully exploit the potential of these materials in their respective applications, these effects need to be independently studied for each material.

Among all the 2D materials, antimonene has high carrier mobility and a suitable band gap, making it viable as a semiconductor for potential applications in electronic, optoelectronic, and even spintronic devices. Moreover, it can be easily synthesized by mechanical exfoliation and cleavage.[27, 28] However, it was only recently discovered in 2015[13] and experimentally produced in 2016,[29] and many of its properties under various complex situations are not fully understood, especially with ripples. Thus, the properties of rippled antimonene warrant further investigation.

The band gap, the main characteristic defining the electrical conductivity of a solid as mentioned above, has been one of the focuses in semiconductor engineering. Many studies have been devoted to tuning the band gap and understanding the mechanisms behind it. Additionally, the work function (WF) is another critical property of a semiconductor. For the normal functioning of piezoelectric devices made from semiconducting 2D materials, the electrical contact should possess a Schottky junction that prevents electrons from crossing the interface and hence neutralizing the piezoelectric polarization charges. The manufacture of Schottky contacts relies on the difference in the WFs of the contacting materials. An in-depth understanding of the WF is essential for achieving efficient charge injection and transport across the heterojunctions. Moreover, understanding the piezoelectricity and flexoelectricity of a material could provide more insights into polarization and charge localization of the material, which will be helpful in designing sensors, actuators, triboelectric nanogenerators, *etc*. These properties can be accurately computed using density functional theory (DFT) calculations.[30] As demonstrated above, the band structure (BS), WF, and polarization greatly influence the application of an electronic material and are therefore the focus of this work.

These properties can be moderated by physical and chemical means. Strain engineering is such a strategy, which has been well studied and widely adopted to modulate the band gap and induce phase transitions.[31,32] Epitaxial strain may also be generated spontaneously in thin films by lattice mismatch between the film and its substrate, during film growth, or under thermal expansion. The common existence and utilization of strain make further understanding of the material behavior under strain and mechanisms of such behavior necessary.

Here, the properties including the WF and BS of rippled β-phase antimonene (β-Sb), which is the most stable phase,[33] under compression are studied *via* DFT computations. The results are compared with those of the flat counterparts to reveal the effects of rippling. Uniaxial compression of 2.5%, 5%, and 7.5% along the armchair and zigzag directions is applied to investigate the anisotropy of the effects. Extensive analyses of the density of states (DOS) and structure geometry are performed for both the rippled and flat structures to account for the differences in the results. The trends in the out-of-plane dipole moment of the rippled structures are also presented.

This work goes beyond the current understanding limited to flat antimonene, which is used as the reference to reveal the effects of rippling. The pristine form is physically difficult to achieve because 2D lattices may prefer to have corrugations, and the autonomous form will always contain ripples. Yet, the highly consistent electronic properties of the rippled form across all the compression levels shed light on ways to restore the properties of the pristine form. The implied possibility to maintain constant performance under compression makes antimonene a robust and reliable candidate for relevant electronic applications. The results presented here are highly practical and can be adopted for potential applications in the future.

## II. COMPUTATIONAL METHOD

### A. Supercell construction

Supercells need to be constructed to accommodate corrugations. The β-Sb unit cell structure used for supercell construction [FIG. 1(a)] is obtained from the study conducted by Kripalani *et al.*,[34] where the $x$, $y$, and $z$ axes are along the zigzag, armchair, and normal directions, respectively. The supercells are constructed by arranging 15 unit cells along the investigated direction (armchair: $1 \times 15$/zigzag: $15 \times 1$) as shown in FIG. 1(b). The computational results are cross checked with those obtained using supercells with 10 unit cells. Uniaxial compression of 2.5%, 5%, and 7.5% is applied along the two directions, resulting in different lattice constants of the supercell along the compressed direction. It is to be highlighted that the compression values mentioned here refer to the contraction of the supercell size, which implies the deformation of the overall configuration. These values are to be distinguished from the local strain, which may vary at different locations, as demonstrated in Section III-C.

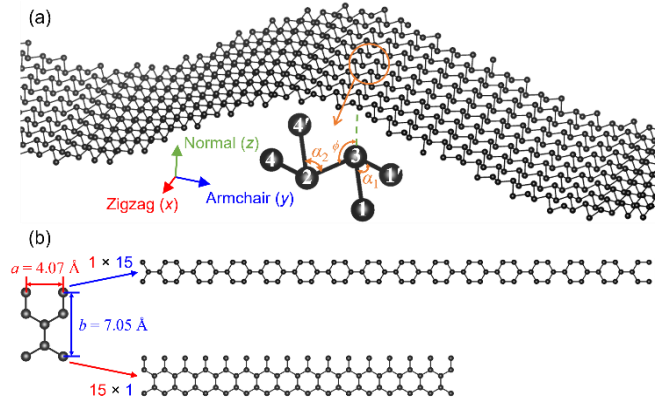

FIG. 1: (a) Rippled β-Sb structure with a zoomed-in schematic illustration of a unit cell (4 atoms) and atoms in the neighboring unit cells (labeled in prime) for showing bond angles $\alpha_1$, $\alpha_2$, and $\phi$. The green dashed line indicates the normal direction. (b) Supercell construction.

B. Relaxation

With zero magnetic moments obtained in trial calculations, non-spin-polarized DFT[35-37] calculations are conducted. The Vienna Ab initio Simulation Package[38] is used with the Perdew–Burke–Ernzerhof exchange–correlation functional[39] under the generalized gradient approximation functional methods. The DFT-D2 method of Grimme[40] is used to correct for van der Waals forces. A kinetic energy cutoff of 450 eV is chosen for the plane wave basis set. Using the Monkhorst–Pack method, 12 $k$ points are sampled in the transverse direction, while 1 $k$ point is used in the other directions in the Brillouin zone. Periodic boundary conditions are applied in the armchair and zigzag directions. Along the normal direction, with a lattice constant of 50 Å, free boundary conditions are enforced with sufficient vacuum separation that eliminates spurious interactions between the rippled slab structures. The energy convergence criterion is set as $10^{-4}$ eV, and all the structures are relaxed until the maximum Hellmann–Feynman force per atom is less than 0.02 eV/Å.

After full relaxation, the following parameters of the pristine structure are obtained: $d_{13} = d_{23} = d_{24} = d_{14'}$ = 2.879 Å, where $d_{ij}$ represents the bond length between atoms $i$ and $j$; $\phi = 125.31°$ and $\alpha_1 = \alpha_2 = 89.94°$, which are consistent with values in the literature.[13, 33, 34, 41] The structure with the lowest energy at each compression level is determined to be the optimized rippled structure. To obtain this structure, a scanning is performed at each compression level along the investigated directions by introducing different initial sinusoidal out-of-plane displacements.

Lattice constant optimization in the transverse (non-loaded) direction is performed using the $1 \times 15$ and $15 \times 1$ flat structures under 7.5% compression, where similarly, the structure with the lowest energy is considered to possess the most stable configuration. Only ~0.7% increase is observed in the lattice constant along both the zigzag and armchair directions. The small increase indicates that the Poisson effect can be ignored for the small compression applied, and the transverse lattice constants can be kept constant for the different compression levels investigated in this study.

### C. Electronic property calculation

It is to be highlighted that this work is to examine the effect of rippling qualitatively. With the relatively large supercell sizes and great total number of cases covered, calculations without considering spin–orbit coupling are adopted to reduce the calculation time, which are also sufficient for observing the general trend and effects.

For electronic property calculations, the energy convergence criterion and maximum Hellmann–Feynman force per atom are reduced to $10^{-6}$ eV and 0.01 eV/Å. The DOS, WF, and BS are computed for each optimized structure. The high symmetry path S–X–Γ–S–Y–Γ for an orthorhombic cell is used for the BS computation, where 20 $k$ points are sampled in each subpath (100 in total). The consistent results obtained by calculations using a single unit cell and the pristine supercells prove that the calculations using the supercells are as reliable.

### D. Polarization

The out-of-plane polarization along the $z$ direction is calculated for the rippled structures using the method adopted by Tan *et al.*[42] The born effective charge tensor $Z_{ij}*$ of each atom is determined from the response to a finite electric field with a strength of 0.01 eV/Å in the $x$, $y$, and $z$ directions. The out-of-plane dipole moment $\partial d_3$ of each atom can be calculated by

$$\partial\boldsymbol{d}_3 = \boldsymbol{Z}^*_{31}\partial\boldsymbol{r}_1 + \boldsymbol{Z}^*_{32}\partial\boldsymbol{r}_2 + \boldsymbol{Z}^*_{33}\partial\boldsymbol{r}_3, \tag{1}$$

where $\partial r_j$ is the displacement of the atom along the $j$ direction from its original position in the flat configuration. The out-of-plane polarization $\boldsymbol{P}_3$ can be determined as

$$\boldsymbol{P}_3 = \frac{\sum_i \partial\boldsymbol{d}_3}{V}, \tag{2}$$

where $V$ is the volume of the involved atoms in the sum. On a per-atom basis, *i.e.*, only one atom is involved, $\boldsymbol{P}_3$ of the atom is just $\partial\boldsymbol{d}_3$ scaled with a constant $V$, which is assumed to be the same for all the atoms.

## III.  RESULTS & DISCUSSION

Consistent results are obtained using supercells with 15 and 10 unit cells, indicating that within the specified range of compression levels, the effects of rippling are independent of the supercell size used for computation, *i.e.*, the wavelength of the corrugations. All the rippled structures exhibit similar properties regardless of the compression level, while obvious trends are observed in their flat counterparts, implying that significant changes are induced by rippling. While this ripple-free case may be nonphysical if the material is in isolation, the ripple deformation mode could in principle be passivated under suitable stacking configurations, for example, *via* strong interlayer/substrate interactions or encapsulation by capping layers. The comparison between the planar and rippled structures under compression reveals not only the effect of rippling but also possible practical implications.

In the following subsections, detailed results on the optimized rippled structures (Subsection A) and the effects of rippling on the electronic properties (Subsection B) are presented for the supercell with 15 unit cells, unless stated otherwise; possible reasons for the similar properties among the rippled structures are deduced from their DOS and structural geometry analyses (Subsection C). The effect of compression on the polarization of the fully relaxed rippled structure is also investigated (Subsection D).

## A. Optimized rippled structure

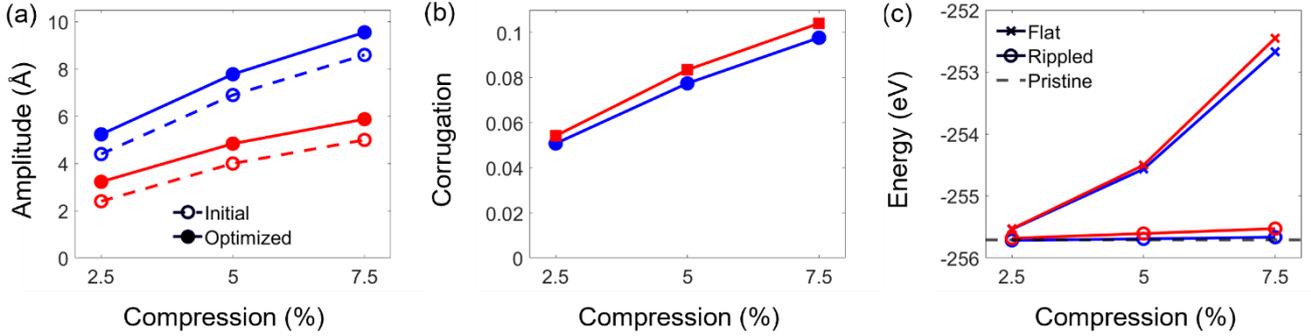

FIG. 2: (a) Amplitude of the initial sinusoidal out-of-plane displacement introduced into the supercell and amplitude of the waveform of the most stable (optimized) rippled structure obtained after DFT relaxation under compression along the armchair and zigzag directions. (b) Corrugation of the optimized rippled structures. (c) Energy of the optimized rippled structures and their flat counterparts under the same compression levels. The energy of the pristine structure is indicated by a gray dashed line for reference. Blue: armchair compression; red: zigzag compression.

As the applied compression increases, the optimized ripple amplitude also increases, as illustrated in FIG. 2(a). As mentioned in Section II-B, sinusoidal out-of-plane displacements are introduced to initiate the rippling. The amplitude $A$ of the optimized structure after full relaxation is constantly larger than that of the sinusoidal waveform before the relaxation, implying that this waveform is not an accurate description of the ripples.

Since the supercells under compression along the two directions have different dimensions, a normalized parameter should be introduced for a fair comparison. The corrugation $C$, defined as $C = A/\lambda$, is plotted in FIG. 2(b) for the optimized rippled structures, where $\lambda$ refers to the wavelength of the ripple. The corrugation value increases with the compression at a decreasing rate, and compression in the zigzag direction induces rippling with a slightly greater corrugation. While rippling in antimonene has never been studied, rippling has been investigated in other 2D materials such as graphene in detail.[43] The corrugation due to thermal fluctuations, edge instabilities, strain, dislocations, *etc*., could vary between 0.05 and as high as 0.5. The corrugation values obtained in our work is at the lower end of this range.

FIG. 2(c) shows that all the rippled structures have lower energies than their flat counterparts at the same compression levels, indicating that rippling gives a more stable structure under compression and hence should be preferred in nature, which is consistent with the Mermin–Wagner theorem. As the compression increases, the energy of the flat structure increases, but that of the rippled structure remains relatively constant and is almost the same as that of the pristine structure. It is therefore evident that under compression, rippling deformation offers a highly favorable pathway to minimize the strain energy and retain the stability of the nanostructure, at least within the compression range of this study. Note that the deviation of the energy of the rippled structure from that of the pristine structure is slightly greater under compression along the zigzag direction, and the deviation increases with the compression.

## B. Electronic properties

The electronic properties of a material determine its applications. Herein, the WF and BS of both rippled and flat antimonene structures under compression are studied and compared, and the DOS is plotted to rationalize their trends.

### 1. Work function

As the compression increases, the WF of the flat structure decreases as shown in FIG. 3(a), indicating that less energy is needed to remove an electron from the structure to a point in vacuum immediately outside its surface. A greater change is experienced by the structure under zigzag compression rather than under armchair compression. On the other hand, the WF of the rippled structure remains relatively constant and deviates only slightly from the value of the pristine structure. The simulation results also show that deviation from this constant value could occur for non-optimized structures. It can therefore be inferred that for applications where the structure may undergo compressive strains in any direction, rippled antimonene would be a suitable candidate to prevent fluctuations in charge transport across its interface as long as it is allowed to fully relax.

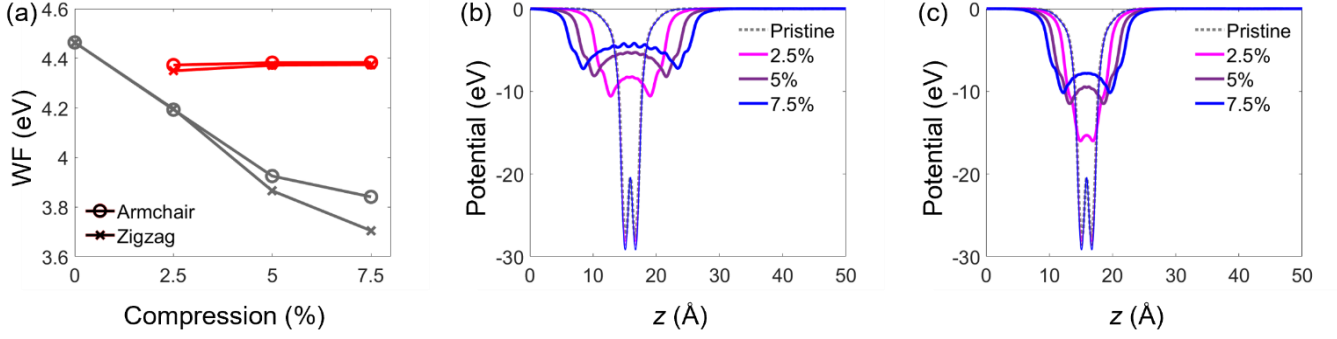

FIG. 3: (a) WF of both the rippled (red) and flat (gray) supercell structures with 15 unit cells under the same compression levels. Average potential in the *x-y* plane of the rippled (thick lines) and flat (thin lines) structures along the *z* direction under (b) armchair and (c) zigzag compression. The vacuum energy level is calibrated to be 0 eV for all.

Significant changes in the plot of the average potential in the *x-y* plane along the *z* direction are induced by rippling, as shown in FIG. 3(b and c). Different from the flat structure where the two sublayers possess only two different *z* coordinates, the rippled structure has atoms with a range of *z* coordinates, thereby widening the dip and decreasing the depth of the potential well. The flat potential from $z = 30$ Å onwards indicates that the selected lattice constant of 50 Å along the normal direction is big enough to eliminate spurious interactions between the rippled slab structures under periodic boundary conditions.

*2. Band structure*

Pristine antimonene is a semiconductor with an indirect band gap. As shown in FIG. 4(a), its conduction band minimum (CBM) is along the Y–Γ path, while the valence band maximum (VBM) is at the Γ point. The band gap (CBM – VBM) computed for the unit cell and pristine supercells are both 1.13 eV, consistent with the results obtained by Kripalani *et al.*[34] Depicted in FIG. 4(b and c), as the compression increases, the band gap of the flat structure narrows down. The reduction in the band gap is mainly due to the decrease in the CBM. When the armchair compression reaches 7.5%, the band gap closes, which indicates a semiconductor–metal transition. Such a transition does not occur in the zigzag-compressed flat structure within the compression range investigated. In contrast, the rippled structure still exhibits a finite band gap, which is of relatively constant value at all the compression levels. The electronic properties at other ripple amplitudes are also calculated. Similarly to the case of WF, a small deviation in the amplitude from the optimized value could result in a big change in the band gap or even shift the CBM position.

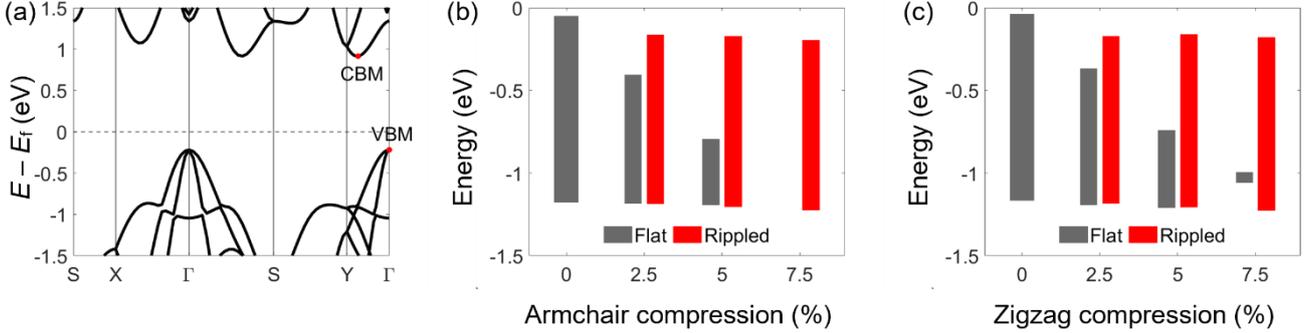

FIG. 4: (a) BS of pristine antimonene (unit cell model). The Fermi level is represented by $E_\text{f}$, and the CBM and VBM are labeled accordingly. Variation of the CBM (top), VBM (bottom), and band gap (bar) of the flat (gray) and rippled (red) structures with compression along the (b) armchair and (c) zigzag directions with calibration of the vacuum energy level to 0 eV. There is no band gap under 7.5% armchair compression for the flat structure because of semiconductor–metal transition. All the supercells involved contain 15 unit cells.

The highly stable WF and band gap of rippled antimonene under compression may find their applications in strain engineering. From this study, it becomes clear that rippled antimonene needs to be fully relaxed to maintain its electronic properties and performance under complex working environments where the structure is prone to strains. Meanwhile, the results highlight the interesting possibility of tuning the WF and band gap of antimonene across a big range of values simply by controlling the corrugation of the 2D material.

*3. Density of states*

To investigate the orbital contribution in the BS and explain the mechanism behind the relatively constant electronic properties, the total and projected DOS are plotted. As shown in FIG. 5(a and c), the p orbital states dominate the regions around the CBM and VBM. They are highly delocalized near the VBM, implying that antimonene possesses exceptional carrier mobility and great transport efficiency. The contribution of the s orbital states increases at the low-energy region of the valence band, and the contribution of both the s and d orbitals increases at the high-energy region of the conduction band, indicating the presence of orbital hybridization at these energy levels. The DOS in FIG. 5(b) are continuous because antimonene becomes a metal in this case.

The contribution of each p orbital ($p_x$, $p_y$, and $p_z$) is further investigated [FIG. 5(d–f)]. From the comparison between the DOS of the rippled and flat structures under armchair compression, it can be inferred that for the flat structure, the increase in compression mainly causes a downward shift of the energy of $p_y$ and $p_z$ orbitals in the conduction band with reference to the Fermi level, which eventually leads to the closing of the band gap at 7.5% compression. The $p_x$ orbital is less affected because the compression is applied along the $y$ direction. Similarly, the energy of the $p_y$ orbital shifts less than that of the other p orbitals under zigzag compression. This energy shift could be attributed to the increase in overlap between the orbitals of neighboring atoms under strains. On the other hand, the rippled structures maintain the energy of these orbitals and have the same DOS pattern at all the compression levels along both directions.

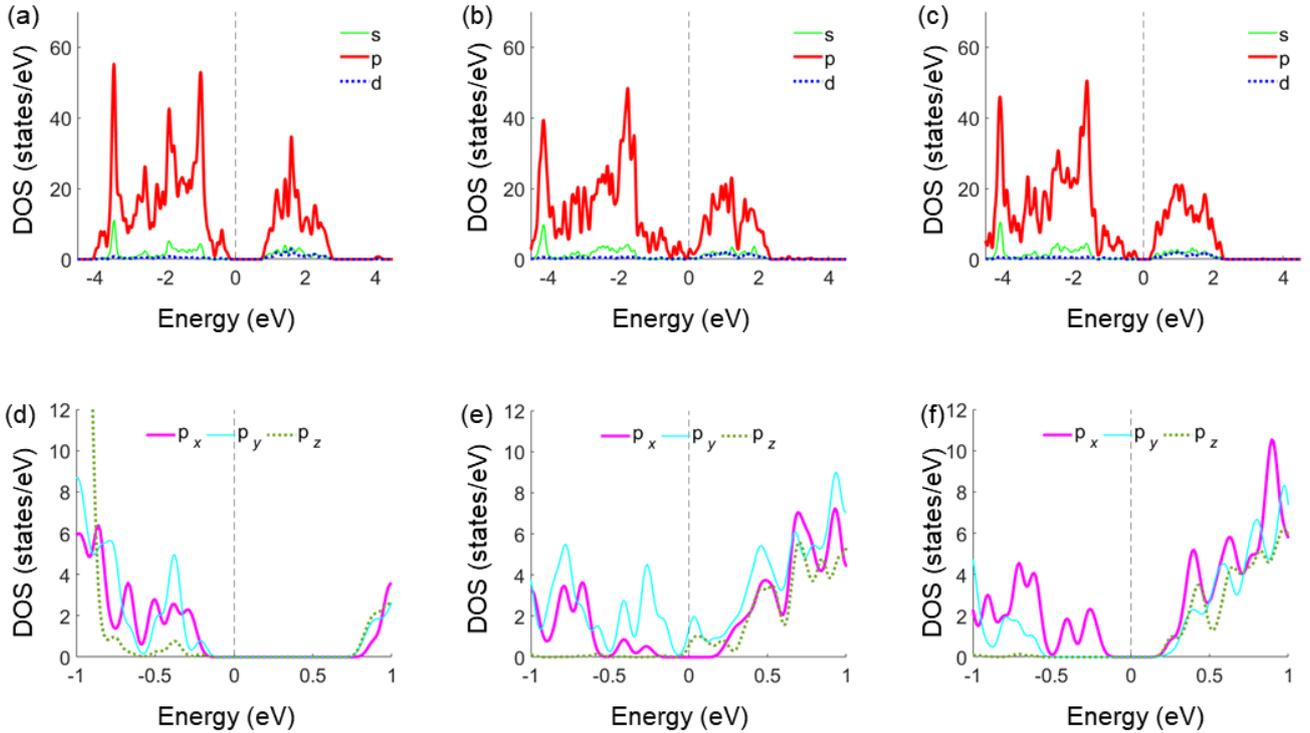

FIG. 5: DOS of the (a) rippled and (b) flat structures under 7.5% armchair compression and (c) DOS of the flat structure under 5.0% zigzag compression. (d–f) Respective zoomed-in plots for the p orbitals around the Fermi level.

## C. Geometry analysis

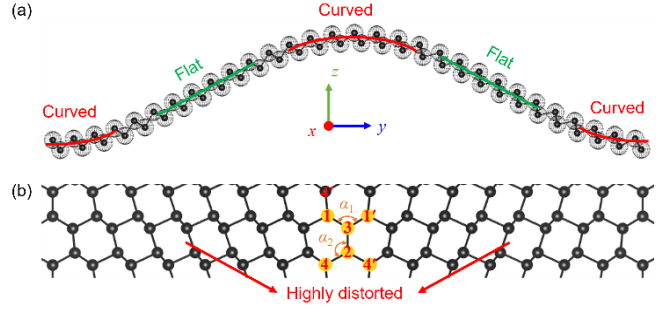

FIG. 6: (a) Rippled structure under 7.5% compression along the armchair direction. (b) Top view showing the configuration of the distorted optimized rippled structure under 7.5% zigzag compression, where the part highlighted in yellow (peak) and the end (trough) are the least distorted regions.

To understand the energetics of β-Sb under compression and rationalize the highly stable electronic properties of the rippled structures, the geometric characteristics of the orthorhombic cell during structural deformation are investigated in detail. As shown in FIG. 6(a), the rippled structure is curved at the peak and trough but rather flat in between.

The bond lengths and angles define the configuration of the 2D lattice of antimonene. For the supercell with 15 unit cells under low armchair compression [FIG. 7(d)], the variations in the bond lengths are rather symmetric with slightly greater values in the flat regions of the structure. The rest of the cases under armchair compression [FIG. 7(a–c, e, and f)] exhibit asymmetric variations: the left and right sides deviate in opposite directions from the average value, especially in the supercells with 10 unit cells. However, the scale of such variations is small compared to the magnitude of the bond lengths, especially for $d_{13}$ and $d_{24}$. The small deviation and fluctuation of these two zigzag bonds may be attributed to the bulky configuration around them as shown in FIG. 6(a).

While the bond lengths exhibit different variations, the bond angle changes under armchair compression are more regular, as shown by FIG. 7(g–l). The bond angle $\alpha_1$ remains constant because there is no transverse movement of the atoms.

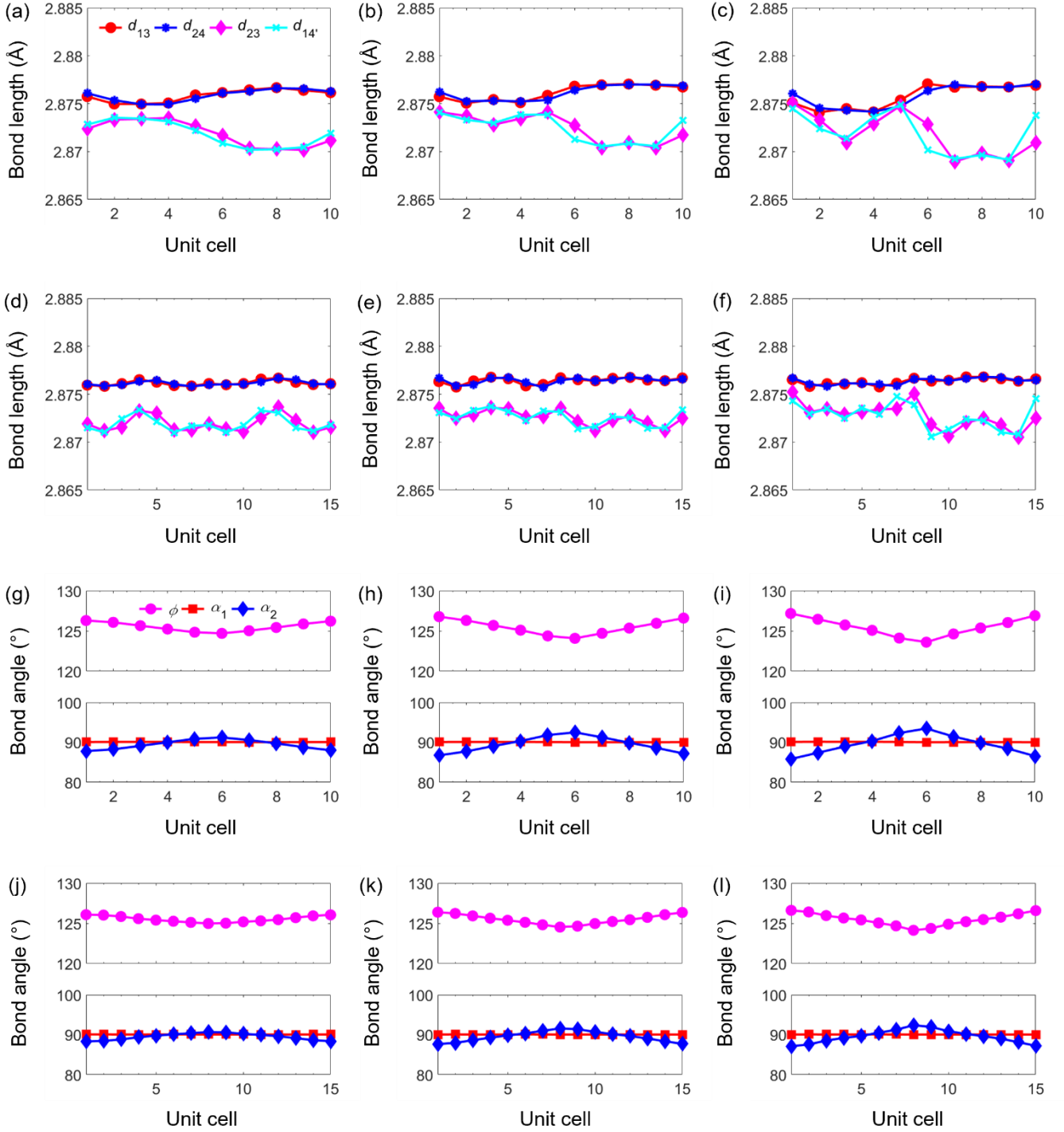

FIG. 7: (a–f) Bond lengths and (g–l) bond angles of the optimized rippled structures with 10 (a–c and g–i) and 15 (d–f and j–l) unit cells in the supercell under compression of 2.5% (a, d, g, and j), 5% (b, e, h, and k), and 7.5% (c, f, i, and l) along the armchair direction.

The structures under zigzag compression exhibit similar behaviors. The shape of the originally buckled honeycomb structure is highly distorted, except at the curved regions of the wave, as shown in FIG. 6(b). Atoms 2 and 3 form bonds with atoms 4 and 1, respectively, on both sides. The variation of all the bond lengths across the optimized rippled structures are depicted in FIG. 8(a–f). More symmetrical results are obtained for supercells with both 10 and 15 unit cells under zigzag compression than under armchair compression. The lengths of the zigzag bonds ($d_{13}$ and $d_{24}$) experience greater fluctuations, while the armchair bonds ($d_{23}$ and $d_{14'}$) are relatively constant and generally longer because they are less affected by the zigzag strain. The exception is in the supercell with 10 unit cells under 7.5% compression, where the armchair bonds are shorter in the curved region of the wave. The reason could be that the length of the supercell along the zigzag direction is insufficient to accommodate the high corrugation under the high compression. The difference between the zigzag and armchair bond lengths decreases as the compression increases. There is high overlap between the variations of $d_{13}$ and $d_{24'}$ as well as $d_{31'}$ and $d_{24}$. It could be due to the similar force fields experienced by the parallel bonds in proximity. All the bond lengths are directly related to the curvature of the structure.

Different from the case of armchair compression, all the bond angles experience significant variations across the structure under zigzag compression, as shown in FIG. 8(g–l). The bond angle $\alpha_1$ decreases at the peak of the structure because the atoms along the $x$ axis (atom 1 in each unit cell), being located in the bottom sublayer of the structure, are compressed together. The distortion of the structure is reflected by the variation of $α_2$ as the parallel bonds fluctuate together. The angle $ϕ$ has similar variations to the case of deformation under armchair compression.

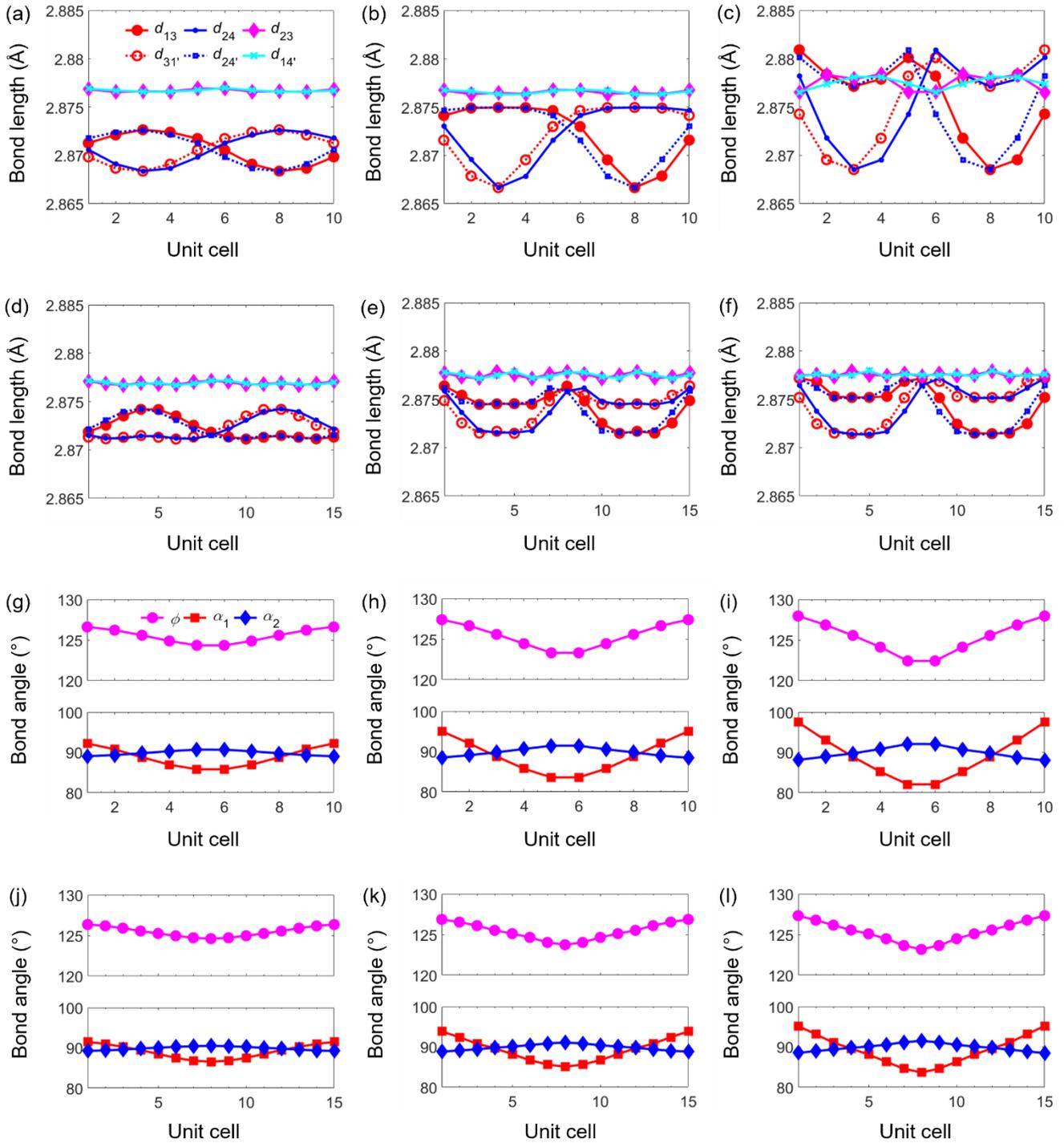

FIG. 8: (a–f) Bond lengths and (g–l) bond angles of the optimized rippled structures with 10 (a–c and g–i) and 15 (d–f and j–l) unit cells in the supercell under compression of 2.5% (a, d, g, and j), 5% (b, e, h, and k), and 7.5% (c, f, i, and l) along the zigzag direction.

After observing the variations of the bond lengths and angles along the compression direction, it is important to see the overall effect of rippling by comparing the mean values to those of the flat counterparts. As shown in FIG. 9(a and b), the mean bond lengths of the rippled structure (15 unit cells) remain relatively constant under different compression levels and are only slightly shorter than those of the pristine structure. This negligible deviation from the case of the pristine structure may be attributed to the relaxation that allows the bonds to be stretched overall to release the strain energy. In contrast, the flat structures are unable to reorientate the atoms, leading to more compact configurations in general as demonstrated by the much shorter mean bond lengths, which have lower values as the compression increases. For the same structure, the zigzag bonds $d_{13}$ and $d_{24}$ have the same mean value, and the armchair bonds $d_{23}$ and $d_{14'}$ have the same mean value. The zigzag bonds are on average less compressed than the armchair bonds under armchair compression and more compressed than the armchair bonds under zigzag compression.

Similarly, although the bond angles exhibit variations at different curvatures of the rippled structure, the mean values across the supercell remain relatively constant under different compression levels and are almost the same as those of the pristine structure, as shown in FIG. 9(c and d). In contrast, because the configuration of the flat structure becomes more compact when the compression increases, $\phi$ increases under both the armchair and zigzag compression. The direction of the compression determines the changes in $\alpha_1$ and $\alpha_2$. The angle $\alpha_1$ increases slightly under armchair compression but decreases under zigzag compression, while $\alpha_2$ decreases in both cases with a greater magnitude under armchair compression.

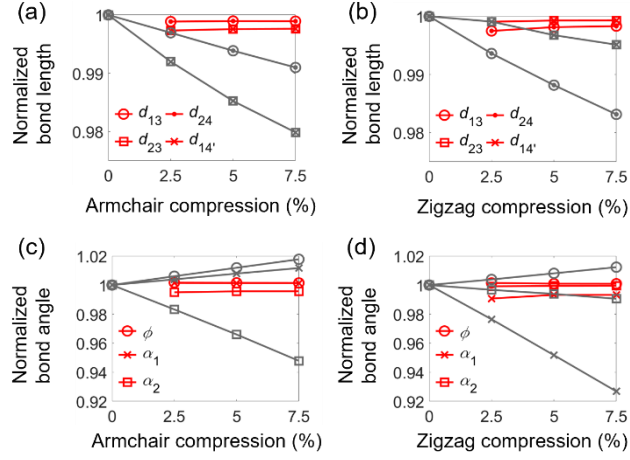

FIG. 9: Mean bond lengths (a and b) and mean bond angles (c and d) of the optimized rippled structures (red) and their flat counterparts (gray) under armchair (a and c) and zigzag (b and d) compression, normalized with the values of the pristine structure. The mean value of $d_{13}$ and $d_{31'}$ ($d_{24}$ and $d_{24'}$) is plotted as $d_{13}$ ($d_{24}$) for the zigzag compression.

To characterize the variations in the bond lengths and angles across the wave, discrete geometry analysis (DGA) [44] is performed. DGA is a well-established way to determine strain in 2D materials, and it has been applied by Kistanov *et al.* [45] to determine the strain induced by the cross-sheet motion of lithium through antimonene sublayers. In our work, using the method of triangulation over a finite mesh of atomic positions, the metric tensor $g$, mean curvature $H$, and Gaussian curvature $K$ for each of the two sublayers of the rippled antimonene structure are computed. The invariants Tr($g$) and Det($g$) represent the local strain of the material: they take on values greater than 1 under in-plane tension but less than 1 under in-plane compression. The curvature values represent the out-of-plane deviation from planarity and character of the surface profile ($H = K = 0$ for planar configuration, $K > 0$ for elliptical configuration, and $K < 0$ for hyperbolic configuration).

As shown in FIG. 10, the top and bottom sublayers of the rippled structure exhibit opposite changes in both Tr($g$) and Det($g$), which proves the existence of local strain within a single sublayer, although the average values are around 1. A similar trend is also observed in rippled structures at other compression levels, which explains the consistent electronic properties of the optimized rippled structures regardless of the compression level. In contrast, the flat structure under the same compression level experiences high, uniform compressive strain. Compared to the flat structures (the flat structure under 7.5% armchair compression and pristine structure under no strain), the rippled structure exhibits hyperbolic configuration for its both sublayers. The significant, nonuniform curvature exemplifies structural deformation from the pristine structure. The unity value obtained for Tr($g$) and Det($g$) of the pristine structure and zero value obtained for the curvature of the flat structures proves the validity of the calculations.

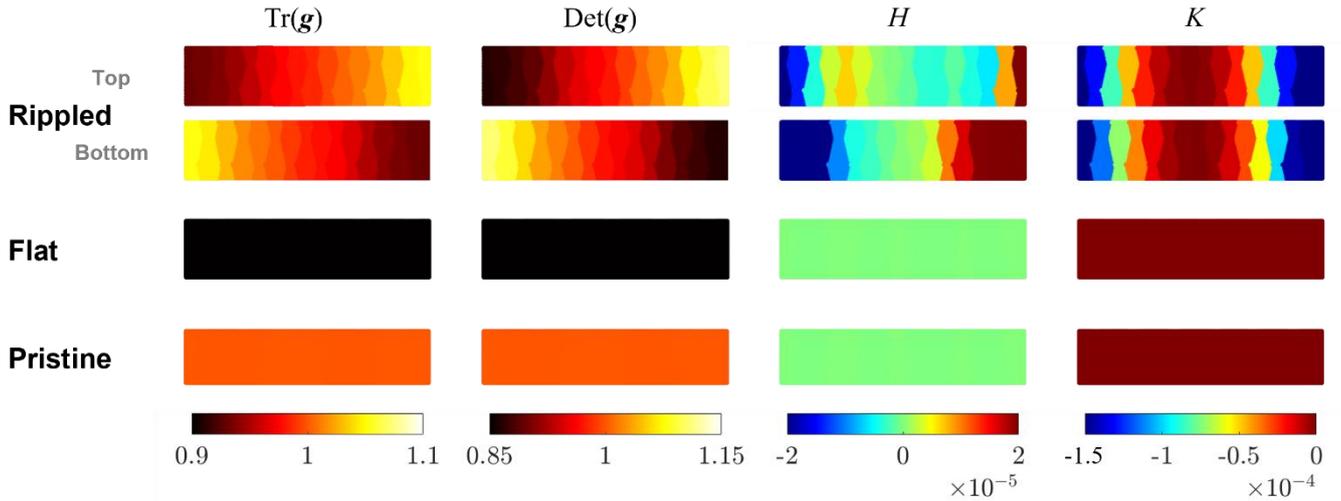

FIG. 10: DGA of rippled (top and bottom sublayers) and flat antimonene under 7.5% armchair compression and pristine antimonene (strain free). Since the same results are obtained for both sublayers of the flat and pristine forms, only those of the bottom sublayers are presented.

It can be concluded that by introducing ripples and allowing full relaxation of the antimonene structure, the configuration of antimonene can be preserved overall under compression, as long as the wavelength of the ripple is long enough to accommodate the corrugation. Under uniaxial compression along both the armchair and zigzag directions, antimonene is able to release the strain energy through rippling, unlike phosphorene, where ripple deformation under compression occurs only along the zigzag direction.[46] This behavior may make tuning of antimonene easier because the strain direction is not a concern. The relatively constant configuration of the rippled structure proves that the overlap between the p orbitals of the neighboring atoms is less altered on average, which helps to retain the electronic properties of the pristine structure.

### D. Piezoelectricity

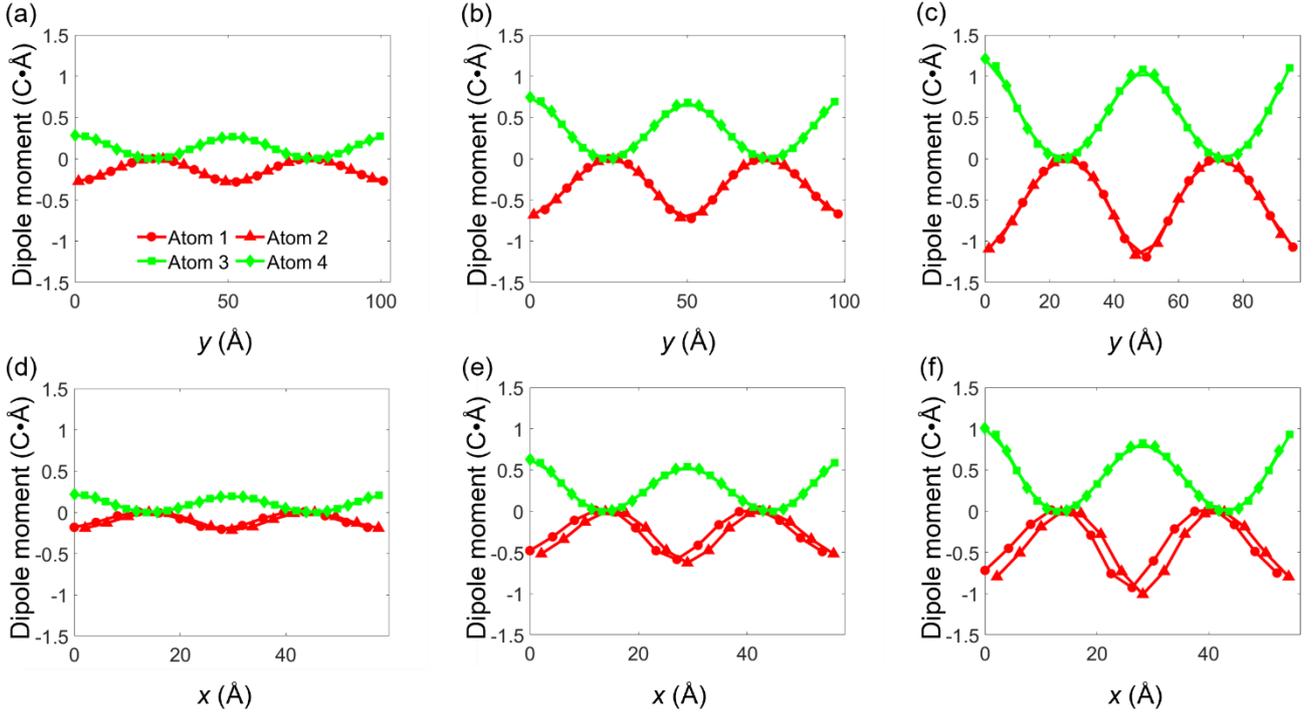

FIG. 11: Out-of-plane dipole moment of the optimized rippled structures with 15 unit cells under (a–c) armchair and (d–f) zigzag compression of 2.5% (a and d), 5% (b and e), and 7.5% (c and f).

To explore the effect of rippling on the piezoelectric properties of β-Sb, the out-of-plane dipole moment is computed according to eq.(1) and plotted for the rippled structures under compression (FIG. 11). The moment values approach zero where the antimonene structure is rather flat and reach a maximum where it has the maximum curvature. As the compression increases along the armchair and zigzag directions, the magnitude of the dipole moment also increases proportionally with similar values in both cases, which may potentially serve as a driving mechanism for contact electrification.

Different from the case of graphene,[42] of which the structure is perfectly planar and only one atom thick, the moment plots of buckled antimonene have two curves, which show obvious symmetry with atoms 1 and 2 having the negative values and atoms 3 and 4 having the positive values. Note that atoms 1 and 2 are in the bottom sublayer of the antimonene structure, while atoms 3 and 4 are in the top sublayer, which could be the reason for the opposite moment values upon bending of the structure. According to eq.(2), for a free-standing rippled monolayer, these values may be canceled out and give rise to zero polarization overall.

Nevertheless, in antimonene multilayers and its heterostructures, polarization changes induced through corrugations may still be utilized for charge transfer and contact electrification. It has been found through both simulations[42] and experiments[47, 48] that the rippling of materials in contact can enhance charge transfer and thus cause depletion and accumulation of electrons near the concave and convex surfaces, respectively. Electrostatic potential differences can therefore be induced across the interfaces. Having a low electron affinity[49] and great number of electrons, antimonene exhibits high transport efficiency. Antimonene-based materials could hence be a suitable candidate for piezoelectric and flexoelectric devices.

## IV. CONCLUSION

The electronic properties including the WF and BS of rippled β-Sb under uniaxial compression along both the armchair and zigzag directions have been investigated and compared with those of the flat counterparts to reveal the effects of rippling. It has been found that under such compression, the rippling deformation mode is energetically favored over the flat structure along both directions. The introduction of rippling preserves the properties of the pristine antimonene structure (flat and strain free), and these properties do not vary with the compression level applied, at least within the range examined in this study (below 7.5%). The underlying mechanism for this behavior is explained by the orbital contribution based on analyses of the DOS and structure configuration. It is deduced that full relaxation of the rippled structure under compression allows the atoms to rearrange and stabilize to release the strain energy. The relatively constant mean bond lengths and angles of the optimized structures under compression minimize overall changes in the overlap between the p orbitals of the neighboring atoms, which in turn helps to maintain the electronic properties. The extensive computation and thorough analysis highlight the rich strain–amplitude configuration space of rippled antimonene, which can be exploited to tune its electronic properties for advanced functional applications. Additionally, it has been shown that the curvature of the structure and its double-sublayer layout play an important role in its piezoelectric properties.

While the effect of rippling on the electronic properties of antimonene has been demonstrated, the impact on the mechanical or even chemical properties can be further investigated in the future. Dynamic analysis on spontaneous rippling may also be performed to understand the deformation mechanism. Beyond rippling, other factors such as doping, defects, and number of layers may also alter the properties and remain to be studied in future investigations. This work also provides compelling theoretical guidance for the investigation of rippling phenomena in other 2D materials, for which the strain–amplitude–property relationship, to the best of our knowledge, has never been comprehensively examined.

## V. ACKNOWLEDGEMENTS

This study was supported by the Economic Development Board, Singapore and Infineon Technologies Asia Pacific Pte Ltd through the Industrial Postgraduate Program with Nanyang Technological University. The computational calculations for this study were partially conducted using the resources of the National Supercomputing Centre, Singapore.

## VI. REFERENCES

[1]     R. Peierls, Helv. Phys. Acta **7**, 81 (1934).
[2]     N. D. Mermin and H. Wagner, Physical Review Letters **17**, 1133 (1966).
[3]     L. Landau and E. Lifshitz, *Statistical Physics* (Pergamon, Oxford, 1980).
[4]     K. S. Novoselov, A. K. Geim, S. V. Morozov, D.-e. Jiang, Y. Zhang, S. V. Dubonos, I. V. Grigorieva, and A. A. Firsov, Science **306**, 666 (2004).
[5]     T. Lei, J.-M. Li, F.-S. Li, J.-O. Wang, K. Ibrahim, and K. Zhang, Applied Physics Letters **115**, 221602 (2019).
[6]     M. Xie, S. Zhang, B. Cai, Y. Zou, and H. Zeng, RSC Advances **6**, 14620 (2016).
[7]     H. V. Phuc, N. N. Hieu, B. D. Hoi, L. T. T. Phuong, N. V. Hieu, and C. V. Nguyen, Superlattices and microstructures **112**, 554 (2017).
[8]     M. S. Khan, R. Ratn, and A. Srivastava, Pramana **89**, 9 (2017).
[9]     G. Bian, X. Wang, Y. Liu, T. Miller, and T. C. Chiang, Phys Rev Lett **108**, 176401 (2012).

[10] D. R. Kripalani, Y. Cai, M. Xue, and K. Zhou, Physical Review B **100**, 224107 (2019).
[11] S. Agnihotri, P. Rastogi, Y. S. Chauhan, A. Agarwal, and S. Bhowmick, The Journal of Physical Chemistry C **122**, 5171 (2018).
[12] Y. Zhou, L. Yang, X. Zu, and F. Gao, Nanoscale **8**, 11827 (2016).
[13] S. Zhang, Z. Yan, Y. Li, Z. Chen, and H. Zeng, Angew Chem Int Ed Engl **54**, 3112 (2015).
[14] Y. Xu, B. Peng, H. Zhang, H. Shao, R. Zhang, and H. Zhu, Annalen der Physik **529**, 1600152 (2017).
[15] S. Zhang, M. Xie, F. Li, Z. Yan, Y. Li, E. Kan, W. Liu, Z. Chen, and H. Zeng, Angew Chem Int Ed Engl **55**, 1666 (2016).
[16] X. Chen, Q. Yang, R. Meng, J. Jiang, Q. Liang, C. Tan, and X. Sun, Journal of Materials Chemistry C **4**, 5434 (2016).
[17] J. Qi, X. Li, X. Qian, and J. Feng, Applied Physics Letters **102**, 173112 (2013).
[18] Q. Yue, J. Kang, Z. Shao, X. Zhang, S. Chang, G. Wang, S. Qin, and J. Li, Physics Letters A **376**, 1166 (2012).
[19] P. Miro, M. Ghorbani-Asl, and T. Heine, Adv Mater **25**, 5473 (2013).
[20] S. Barraza-Lopez, B. M. Fregoso, J. W. Villanova, S. S. P. Parkin, and K. Chang, Reviews of Modern Physics **93**, 011001 (2021).
[21] B. Huang, G. Clark, D. R. Klein, D. MacNeill, E. Navarro-Moratalla, K. L. Seyler, N. Wilson, M. A. McGuire, D. H. Cobden, D. Xiao *et al.*, Nature Nanotechnology **13**, 544 (2018).
[22] S. Barraza-Lopez, F. Xia, W. Zhu, and H. Wang, Journal of Applied Physics **128**, 140401 (2020).
[23] P. C. Hohenberg, Physical Review **158**, 383 (1967).
[24] A. Fasolino, J. H. Los, and M. I. Katsnelson, Nat Mater **6**, 858 (2007).
[25] J. C. Meyer, A. K. Geim, M. I. Katsnelson, K. S. Novoselov, T. J. Booth, and S. Roth, Nature **446**, 60 (2007).
[26] T. T. Cui, J. C. Li, W. Gao, and Q. Jiang, Phys Chem Chem Phys **20**, 2230 (2018).
[27] P. Ares, F. Aguilar-Galindo, D. Rodriguez-San-Miguel, D. A. Aldave, S. Diaz-Tendero, M. Alcami, F. Martin, J. Gomez-Herrero, and F. Zamora, Adv Mater **28**, 6332 (2016).
[28] K. Sugawara, T. Sato, S. Souma, T. Takahashi, M. Arai, and T. Sasaki, Phys Rev Lett **96**, 046411 (2006).
[29] C. Gibaja, D. Rodriguez-San-Miguel, P. Ares, J. Gomez-Herrero, M. Varela, R. Gillen, J. Maultzsch, F. Hauke, A. Hirsch, G. Abellan *et al.*, Angew Chem Int Ed Engl **55**, 14345 (2016).
[30] R. Hinchet, U. Khan, C. Falconi, and S.-W. Kim, Materials Today **21**, 611 (2018).
[31] B. Amorim, A. Cortijo, F. de Juan, A. G. Grushin, F. Guinea, A. Gutiérrez-Rubio, H. Ochoa, V. Parente, R. Roldán, P. San-Jose *et al.*, Physics Reports **617**, 1 (2016).
[32] G. G. Naumis, S. Barraza-Lopez, M. Oliva-Leyva, and H. Terrones, Rep Prog Phys **80**, 096501 (2017).
[33] G. Wang, R. Pandey, and S. P. Karna, ACS Appl Mater Interfaces **7**, 11490 (2015).
[34] D. R. Kripalani, A. A. Kistanov, Y. Cai, M. Xue, and K. Zhou, Physical Review B **98**, 085410 (2018).
[35] W. Kohn and L. J. Sham, Physical Review **140**, A1133 (1965).
[36] R. G. Parr and W. Yang, in *Proceedings of the Third International Congress of Quantum Chemistry* (Springer, Kyoto, 1979), pp. 5.
[37] D. Sholl and J. A. Steckel, *Density functional theory: a practical introduction* (John Wiley & Sons, Hoboken, New Jersey, 2011).
[38] G. Kresse and J. Furthmüller, Physical review B **54**, 11169 (1996).
[39] J. P. Perdew, K. Burke, and M. Ernzerhof, Physical review letters **77**, 3865 (1996).
[40] S. Grimme, J Comput Chem **27**, 1787 (2006).
[41] J. Liang, L. Cheng, J. Zhang, and H. Liu, arXiv preprint, arXiv:1502.01610v2 (2017).

[42] D. Tan, M. Willatzen, and Z. L. Wang, Nano Energy **79**, 105386 (2021).
[43] S. Deng and V. Berry, Materials Today **19**, 197 (2016).
[44] M. Mehboudi, K. Utt, H. Terrones, E. O. Harriss, A. A. Pacheco SanJuan, and S. Barraza-Lopez, Proc Natl Acad Sci U S A **112**, 5888 (2015).
[45] A. A. Kistanov, D. R. Kripalani, Y. Cai, S. V. Dmitriev, K. Zhou, and Y.-W. Zhang, Journal of Materials Chemistry A **7**, 2901 (2019).
[46] L. Kou, Y. Ma, S. C. Smith, and C. Chen, J Phys Chem Lett **6**, 1509 (2015).
[47] C. Xu, B. Zhang, A. C. Wang, H. Zou, G. Liu, W. Ding, C. Wu, M. Ma, P. Feng, Z. Lin *et al.*, ACS Nano **13**, 2034 (2019).
[48] J. Nie, Z. Ren, L. Xu, S. Lin, F. Zhan, X. Chen, and Z. L. Wang, Adv Mater **32**, e1905696 (2020).
[49] A. Kokabi and S. B. Touski, Physica E: Low-dimensional Systems and Nanostructures **124**, 114336 (2020).